\documentclass[onecolumn, superscriptaddress, prmaterials, amsmath,amssymb,noeprint,11pt]{revtex4-2}

\usepackage{hyperref}
\hypersetup{backref,colorlinks=true,allcolors=prussianblue}
\usepackage{xr}
\usepackage{graphicx}
\usepackage[caption=false]{subfig}

\usepackage[table]{xcolor}
\definecolor{prussianblue}{rgb}{0.0, 0.19, 0.33}

\usepackage{setspace}
\usepackage{float}
\usepackage{indentfirst}

\usepackage{tcolorbox}
\usepackage{todonotes}

\def\fig#1{Fig.~\ref{#1}}
\def\eq#1{Eq.~\eqref{eq:#1}}

\usepackage{tikz}
\usetikzlibrary{calc,arrows,shapes,positioning}
\usetikzlibrary{patterns,snakes}
\usetikzlibrary{positioning,decorations.pathreplacing}

\definecolor{sangria}{rgb}{0.57, 0.0, 0.04}
\definecolor{arsenic}{rgb}{0.23, 0.27, 0.29}
\definecolor{prussianblue}{rgb}{0.0, 0.19, 0.33}
\definecolor{phthalogreen}{rgb}{0.07, 0.21, 0.14}
\definecolor{dgreen}{rgb}{0.0, 0.4, 0.2}

\begin{document}

\author{Okan K. Orhan}
\author{Mauricio Ponga}
\affiliation{Department of Mechanical Engineering, University of British Columbia, 2054 - 6250 Applied Science Lane, Vancouver, BC, V6T 1Z4, Canada}

\title{Engineering ultra-strong Mg-Li-Al-based light-weight alloys from first principles}

\begin{abstract}
Light-weight alloys are essential pillars of transportation technologies.
They also play a crucial role to achieve a more green and cost-effective aerospace technologies. 
Magnesium-lithium-aluminum (Mg-Li-Al) alloys are auspicious candidates due to their promising mechanical strengths at low densities.
We herein present a systematic first-principles investigation of the Mg-Li-Al-based alloys to provide insights for designing ultra-strong light-weight alloys. 
Initial analysis indicates that the Mg-Li-Al mixtures are not thermally stabilized into random-solid solutions.
Following this hint, the base-centered cubic (BCC)-based intermetallics of Mg, Li and Al are investigated for their thermal and elastic stabilities.
Three simple figures of merits are used to further assess their mechanical strengths. 
The most-frequently observed intermetallics are used to predict the yield strength of the hetero-structures from the recent experimental works. 
The rule of mixing works reasonable well to predict the mechanical properties of complex structures starting from isolated intermetallics. 
\end{abstract}

\maketitle

\section{Introduction}
Greener and cost-effective aerospace transportation has been an increasing concern to due to the commitment to cut carbon footprint in half by 2050 in civil aviation~\cite{10.3316/informit.042040113574230}.
Furthermore, the rapidly growing military and commercial space exploration requires light materials which can provide protection against high-speed debris and micro-meteoroids collision~\cite{doi.org/10.1002/advs.202070126,jones2018recent}. 
Both objectives highly rely on development of novel light-weight alloys (LWAs) with exceptional physical properties such as high strength, low density, fatigue durability, slow crack propagation,  corrosion resistance and commercial advantages such as high workability, low-cost repairment/replacement and easy recycling/disposal~\cite{ZHU2018103}.

Mg-Li alloys offer a good balance of strength and ductility~\cite{Cain2020}, whilst AL-Li alloys have  superior fracture toughness as they are commonly more brittle~\cite{Rioja2012}. 
Mg-Li-Al alloys open up a large stoichiometric and phase space to explore LWAs with tunable mechanical properties.
Introducing Al into the Mg-Li mixtures with small molar-fractions have been shown to improve specific strength while keeping the solid density low beyond the grain-refinement strengthening  ~\cite{WANG2011523,FEI2015169}. 
Liu \textit{et al.}~\cite{LIU2007499} has found that the  Mg?14Li-1Al [weight \% (wt\%)], namely LA141, has three phases. 
The main phase is the BCC random solid solution (RSS) of Mg and Li, namely the $\beta$ phase. 
The second predominant phase is the HCP RSS of Mg and Li, namely the $\alpha$ phase, which is embedded into the $\beta$ matrix. 
Additionally, they have observed the BCC-based MgLiAl$_2$, congregating in the $\alpha$ precipitates.  
It has been suggested that the $\alpha$ precipitates and MgLiAl$_2$ provides additional strengthening in this alloy. 

Tang \textit{et al.}~\cite{Tang2019} has found that on the Mg-11Li-3Al (wt\%), namely LA113, has a $\alpha+\beta$ duplex structure with the D0$_3$-Mg$_3$Al precipitates, semi-coherently forming in the $\beta$ matrix. 
In this work, the D0$_3$-Mg$_3$Al precipitates has been suggested as the predominant source of hardening, leading a Vickers hardness ($H_\mathrm{V}$) of $\sim 1.18$~GPa at a low density of $\sim 1.4$~g$\cdot$cm$^{-3}$.
On the other hand, Xin \textit{et al.}~\cite{Xineabf3039} has shown that the predominant hardening is due to the spinodal decomposition to the Al-rich zones in Mg-14Li-7Al (wt\%), namely LA147,  at the low ambient temperature. 
The Al-rich zones with a $~4$~nm diffusion zones are aligned in the elastically soft $[1 0 0]$-direction of the $\beta$ phase. 
The spinodal-decomposition hardening leads to a Vickers hardness of $\sim 1.47$~GPa at a low density of $\sim 1.32$~g$\cdot$cm$^{-3}$  in this sample. 
However, they have also noted that the Al-rich zones gradually transition to the D0$_3$-Mg$_3$Al 
precipitates during the thermal aging. 

In their recent experimental work, Li \textit{et al.}~\cite{LI2022161703} has investigated the phase formation and mechanical strength of the \textit{nearly} equi-molar mixture of Mg, Li and Al. 
However, their sample becomes  Mg-35Li-20Al (wt\%) composition after treatment with a density of $1.68$~g$\cdot$cm$^{-3}$.
Similar to the previous work, their X-ray diffraction (XRD) measurement indicates  the  $\alpha+\beta$ duplex structure. 
XRD also predicts  MgLiAl$_2$, LiAl and Li$_2$Mg phases. 
The energy-dispersive X-ray spectroscopy (EDS) shows the Al-rich regions with the fish-bone micro-structures which are possibly embedded on the Mg-rich matrix. 
They have measured an exception ultimate strength of $\sim700$~MPa and a Vickers hardness of $1.68$~GPa. 

Despite the rapidly growing experimental works, the current literature is limited to a handful of specific compositions.
Furthermore, the nominal and actual chemical compositions may substantially differ after various post-treatments, aging, pollutants etc.
With that challenge in mind, the first-principles simulations become crucial to gain insights when designing LWAs. 
In this work, we present a systematic first-principles investigation on the formation and mechanical properties of Mg-Li-Al-based LWAs.
Our primary objective is to determine effects of intermetallic formation on the strength-density relation independently of complex sample preparations in the experimental works.
The preliminary assessment indicates that the Mg-Li-Al mixtures do not form RSSs. 
This early finding and previous experimental works hint the formation of BBC-based intermetallics. 
The thermal and elastic stability of the most-likely BBC-based intermetallics are investigated. 
First-principles figures of merits (FOMs) are used to predict  ductility/brittleness, plasticity, and hardness of intermetallics. 
The rule of mixing is applied to the experimentally suggested mixtures to predict the accessible strength-density regions for the chosen Mg-Li-Al mixtures.

\section{Theoretical and computational methodology}
The Kohn-Sham formalism of density-functional theory (KS-DFT)~\cite{PhysRev.136.B864,PhysRev.140.A1133} using the semi-local exchange-correlation functionals~\cite{PhysRev.140.A1133,PhysRevB.21.5469,PhysRevB.33.8822,PhysRevB.46.6671} is the almost-ubiquitous approach for simulating the ground-state properties of quantum-mechanical systems. 
In this section, we will briefly introduce the theoretical concepts and their computational evaluation starting from the approximate KS-DFT electronic structures. 
Some common concepts are presented in the {\bf Supplementary Material (SM)} to provide further details for the readers.

\subsection{Figure of merit for random solid solution formation}
Determining the likelihood of RSS formation is a quite challenging task as it requires a reasonable representation of statistical averaging of spatial disorderliness. 
The approximate KS-DFT has challenges when dealing spatially complex systems such as occupational disorderliness which requires simulations on a large number of super-cells to achieve a sufficient statistical representation. 
Virtual-crystal approximation (VCA)~\cite{doi:10.1002/andp.19314010507,PhysRevB.61.7877} provides an expedient tool to assess single-phase multi-component solid at the disordered mean-field limit.  
In practice, VCA is an over-simplified approach by substituting the actual atoms with a single virtual atom represented by an linearly averaged pseudo-potential. 
Despite this, it has been shown to predict well equilibrium bulk properties and simple-phase transformations~\cite{10.3389/fmats.2017.00036}.

An expedient FOM for assessing the thermal stability is the he heat of formation ($H_\mathrm{F}$) given by~\cite{ZHANG2009878} 
\begin{align}
H_\mathrm{F}(\alpha,\beta,P) &= E_{\mathrm{Mg}_\alpha \mathrm{Li}_\beta \mathrm{Al}_{\gamma}}^{P} \nonumber \\
& -\left[ \alpha\;  E_\mathrm{Mg}^\mathrm{HCP}+ \beta\; E_\mathrm{Li}^\mathrm{BCC} +\gamma\; E_\mathrm{Al}^\mathrm{FCC} \right],
\end{align}
where $\gamma = 1-\alpha-\beta$. $E_{\mathrm{Mg}_\alpha \mathrm{Li}_\beta \mathrm{Al}_\gamma}^{p}$ is the total ground-state energy of the RSS of Mg$_\alpha$Li$_\beta$Al$_\gamma$ in the phase $P$.
The phase $P$ are chosen to be either HCP or BCC or face-centered cubic (FCC) as they are most likely crystal structures for simple metallic systems. 
$E_\mathrm{Mg}^\mathrm{HCP}$, $E_\mathrm{Li}^\mathrm{BCC}$ and  $E_\mathrm{Al}^\mathrm{FCC}$ are the total ground-state energies of the HCP Mg, BCC Li and FCC Al, respectively.
RSS formation is only likely for $H_\mathrm{F}<0$.

\subsection{Figures of merits for phase stability of intermetallics}
A more robust and comprehensive stability assessment requires to assess thermal and elastic stability. 
A more accurate thermal-stability FOM is the  mixing Gibbs free energy (GFE), given by~\cite{e21010068}
\begin{align}\label{eq:eq2}
G_\mathrm{mix}=G_{A} -\sum_i^M c_i G_i,
\end{align}
where $G_{A}$, and $ G_i$ are the GFE of $M$-component and its $i^\mathrm{th}$ principle element  with a molar fraction of $c_i$ at their equilibrium volumes. 
For thermal stability, the mixing GFE is expected to be negative at a given temperature. 
For a non-magnetic and pristine solid, the GFE of a solid is the sum of the formation enthalpy ($H(V)$),  the electronic ($F_\mathrm{el}(V,T)$)  and vibrational ($F_\mathrm{vib}(V,T)$) Helmholtz free energies (see {\bf SM} for details on how to approximately obtain these terms within the approximate KS-DFT).
The Born-Huang-stability criteria~\cite{Born:224197,PhysRevB.90.224104} provides the \textit{necessary and sufficient} conditions for the elastic stability of a crystal (see {\bf SM} for the reduced Born-Huang-stability criteria for the cubic and hexagonal symmetries).
These criteria is indeed a special case of the thermal stability by ensuring  that the GFE of an unstrained crystal is a minimum compared to its infinitesimally strained structures~\cite{DEMAREST1977281}.

\subsection{Figures of merits for mechanical properties}
Calculating the mechanical properties of hetero-phased solids is a challenging task as there is large number of mechanism contributes to hardening and strength. 
However, the expedient FOMs can be used to assess workability, strength and hardness.
Poisson's ratio and the Pugh ration are commonly used as FOMs for plasticity and ductility, respectively. 
Higher plasticity indicates well-workability, whilst lower ductility hints more mechanical strength.
However, it is important to note that plasticity is prerequisite of ductility~\cite{ZHU201815}.
Hence, a critical balance between them is required for a well-workable and strong material.
A critical minimum-value (CMV) of $0.25$ for Poisson's ration~\cite{doi:10.1080/14786440808520496}, and a CMV of $1.75$ for the Pugh ratio~\cite{BOUCETTA201459}
 are expected for plasticity, and ductility, respectively.  
The final FOM is the Vickers hardness($H_\mathrm{V}$) given by a semi-empirical relation of  $H_\mathrm{V}=0.151 S$~\cite{teter_1998} to the shear modulus ($S$). 
Moreover, $H_\mathrm{V}$ is approximately related to the yield strength ($\sigma_\mathrm{y}$) with the semi-empirical formula $H_\mathrm{V} \approx 3 \sigma_\mathrm{y}$ (in MPa units)~\cite{TIRYAKIOGLU201517}. 

These FOMs are directly related to the bulk and shear modulus which can be obtained via the Voigt-Reuss-Hill (VRH) averaging~\cite{doi:10.1002/andp.18892741206,doi:10.1002/zamm.19290090104,Hill_1952} of the second-order elastic tensor (see {\bf SM} for further details).
The second-elastic tensor can be calculated by performing a series of ground-state simulations on the infinitesimal-strained structures of a fully relaxed crystal, and using strain-stress relation in the linear-elastic region (see {\bf SM} for the VHR averaging for the cubic and hexagonal symmetries).

\subsection{An expedient approximation for mechanical properties of composite structures}
The mechanical properties of a hetero-phased multi-component solid can be approximated by applying the rule of mixing to the bulk and shear modulus of its constituents. 
The approximate upper and lower bounds are given within the Voigt and Reuss averagings~\cite{doi:10.1002/andp.18892741206,doi:10.1002/zamm.19290090104}, respectively, by
\begin{align}\label{eq:eq3}\
\mathbb{V}=\left(\sum_i^M f_{i}\mathbb{V}_{i}\right) \quad \mathrm{and} \quad
\mathbb{R}=\left(\sum_i^M \frac{f_{i}}{\mathbb{R}_{i}}\right)^{-1},
\end{align}
where $\mathbb{V}_{i}$ and $\mathbb{R}_{i}$ are the Voigt and Reuss-averaged bulk or shear modulus of the $i^\mathrm{th}$ constituent with a volume fraction of $f_{i}$.
The Voigt-Reuss-Hill (VHR)~\cite{Hill_1952} is simply the arithmetic average of $\mathbb{V}_{i}$ and $\mathbb{R}_{i}$ (see {\bf SM} for further details).

\section{Results and discussion}
\begin{figure}[t]
\centering

\includegraphics[width=0.75\textwidth]{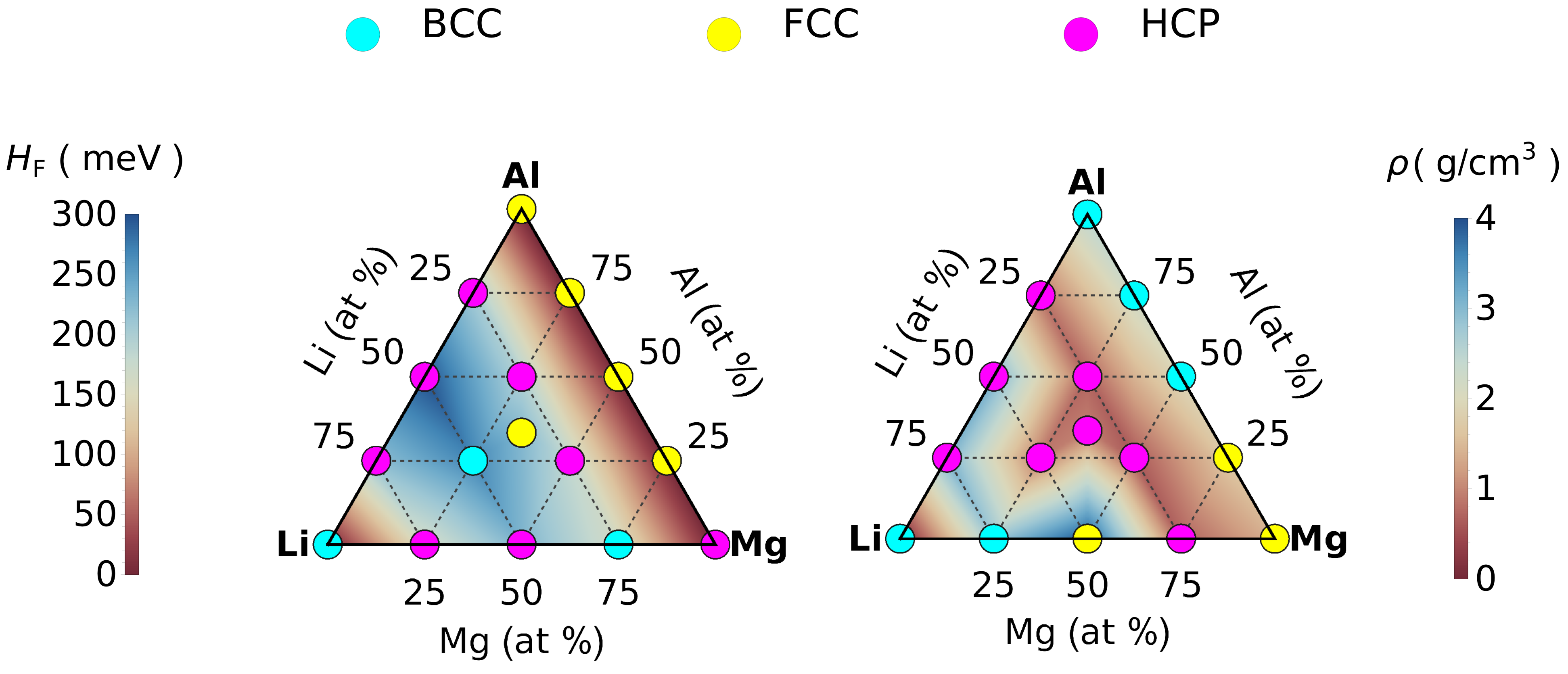}

\caption{Composition-dependence of the  minimum heat of formation ($H_\mathrm{F}$) and minimum solid density ($\rho$) among the BCC, FCC and HCP phases of the Mg$_\beta$Li$_\alpha$Al$_\gamma$ ($\gamma=1-\beta-\alpha$) RSS.
Actual data points are marked by dots colored by their corresponding phases.}
\label{fig:VCA-HoF-Stable}
\end{figure}

We start with the preliminary assessment of the RSS-formation likelihood of Mg$_\beta$Li$_\alpha$Al$_\gamma$ alloys using VCA+KS-DFT.  
In Fig.~\ref{fig:VCA-HoF-Stable} (left), the minimum $H_\mathrm{F}$ among the BCC, FCC and HCP phases of Mg$_\beta$Li$_\alpha$Al$_\gamma$ are shown. 
The positive $H_\mathrm{F}$ throughout the composition space indicates that Mg, Li and Al do not mix into single-phased, spatially disordered multi-component solids. 
Moreover, $H_\mathrm{F}$ are in the order of $10^1-10^2$~meV which is far too high to be possible compensated by the electronic, vibrational and configurational entropic contributions. 
This result can be further verified using the binary bulk metallic model~\cite{doi:10.1080/21663831.2014.985855} calculated via the Miedema's model for pair-wise interaction~\cite{Takeuchi2005}.
In the right panel of Fig.~\ref{fig:VCA-HoF-Stable}, the minimum solid density ($\rho$) among the BCC, FCC and HCP phases of the Mg$_\beta$Li$_\alpha$Al$_\gamma$ are presented. 
The phases with the lowest $\rho$ do not necessarily match with the phase with the lowest $H_\mathrm{F}$. 

\subsection{Body-centered-cubic-based intermetallics}
\begin{figure}[t]
\centering

\includegraphics[width=0.8\textwidth]{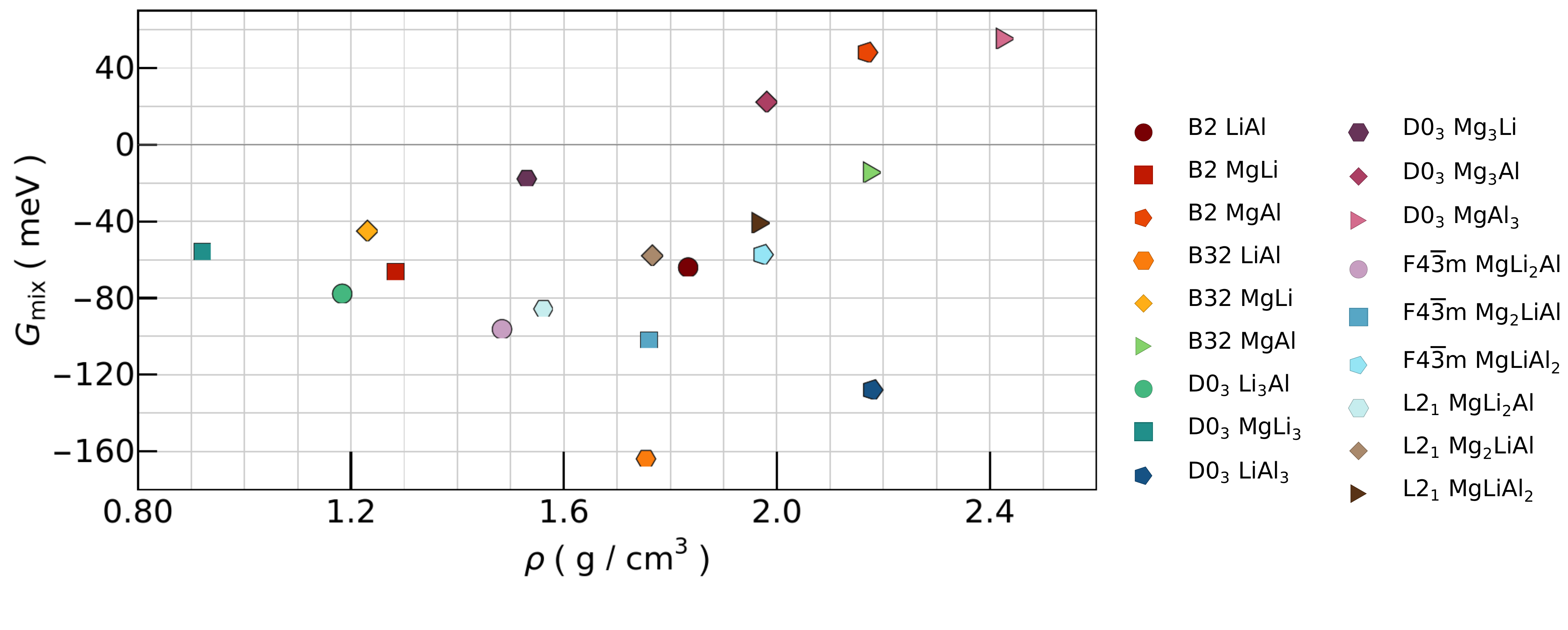}

\caption{Mixing Gibbs free energies of the simple body-centered-cubic-based intermetallics of Mg, Li and Al at $300$~K using the Debye model for the vibrational Helmholtz energies.}
\label{fig:BCC-Gmix}
\end{figure}

Since the Mg$_\beta$Li$_\alpha$Al$_\gamma$ mixtures are not expected to form RSSs, we shift our attention to their candidate BCC-based intermetallic (see {\bf SM} for the templates of the most common BCC-based intermetallics). 
The BCC-based intermetallics are chosen as they are most commonly observed phases and most-likely candidates for high strength~\cite{Tang2019}.
In Fig.~\ref{fig:BCC-Gmix}, the mixing GFE of the isolated intermetallics are presented at $300$~K. 
The mixing enthalpy, $\Delta H$ is equal to $H_\mathrm{F}$ by definition.
$H_\mathrm{V}$ and the mixing vibrational Helmholtz energies ($\Delta F_\mathrm{vib}$) are the predominant terms in \eq{eq2}, whilst the electronic Helmholtz energies ($\Delta F_\mathrm{el}$) are negligible (see {\bf SM-\fig{fig:BCC-HoF}} for the individual contributions of each term). 

In \fig{fig:BCC-Gmix}, the majority of the proposed intermetallics are thermally stable at the room temperature except B2 MgAl, D0$_3$ Mg$_3$Al and  D0$_3$ MgAl$_3$. 
The B32 MgAl and D0$_3$ LiAl$_3$ phases are thermally stabilized by $\Delta F_\mathrm{vib}$, compensating their positive $H_\mathrm{F}$.
$\Delta F_\mathrm{el}$ of the intermetallics mostly work against the thermal stability; however, they are negligible compared to $H_\mathrm{F}$ and $\Delta F_\mathrm{vib}$. 
In general, $\Delta F_\mathrm{vib}$ works against thermal stability at lower densities whilst they improve thermal stability at higher densities. 
Among the thermally unstable phases, the D0$_3$ MgAl$_3$ need further attention due to its particular importance. 
Although the isolated D0$_3$ Mg$_3$Al phase is not thermally stable, Tang \textit{et al.} has shown that it reaches to a meta-stable phase due to its semi-coherent precipitation with the BCC Li matrix~\cite{Tang2019}.

\begin{figure*}[t]
\centering

\includegraphics[width=1.0\textwidth]{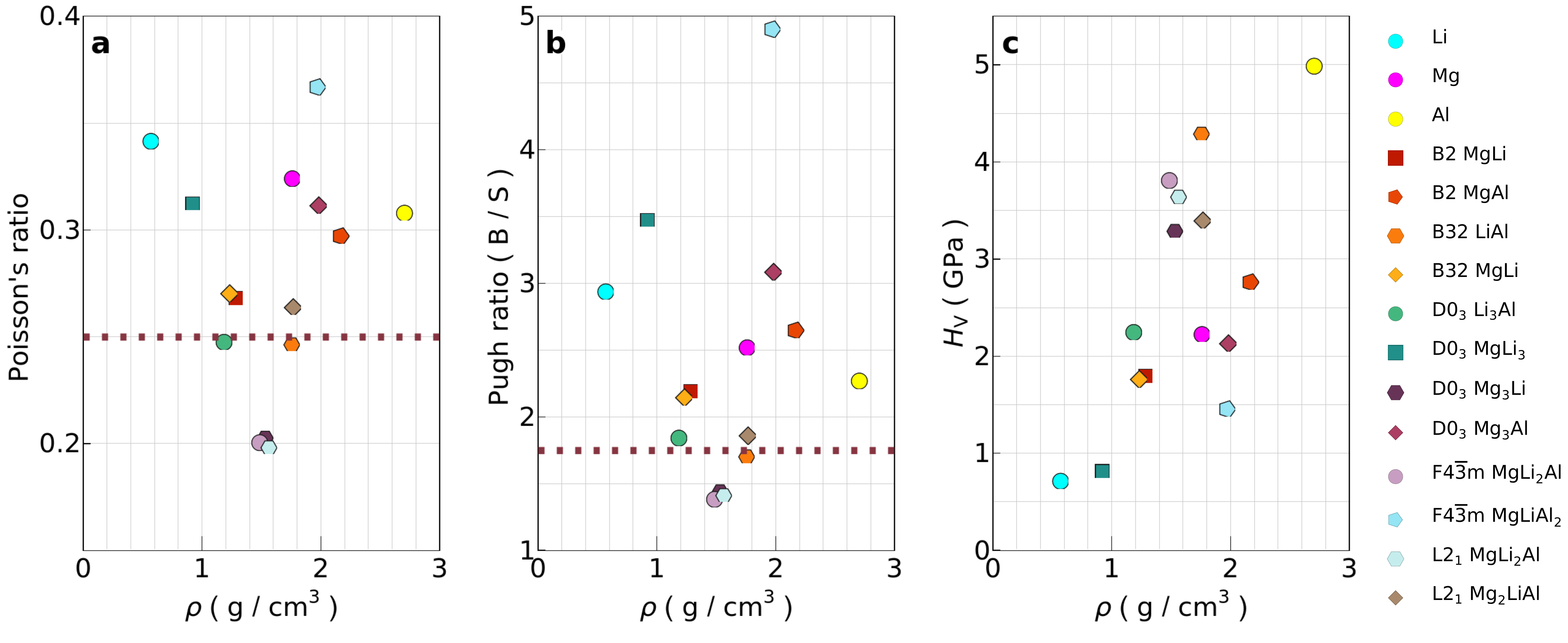}

\caption{Poisson's ratio, Pugh ratio and Vickers hardness of the elastically stable intermetallics of Mg, Li and Al. 
}
\label{fig:BCC-Mechanical}
\end{figure*}

In \fig{fig:BCC-Mechanical}, the three FOMs for the critical mechanical properties are shown for the elastically stable intermetallics alongside the elemental cases. 
The elemental cases behave as expected for which the BCC Li exhibits the lowest strength and hardness with the  best workability and the FCC Al exhibits the highest strength and hardness with the worst workability among Mg, Li and Al.
MgLiAl$_2$ has the highest Poisson's and Pugh ratios, indicating the lowest strength among the available intermetallics. 
This contradicts with the prediction of Liu \textit{et al.}~\cite{LIU2007499} where they have claimed that MgLiAl$_2$ can be an additional source of hardening.
Moreover, the  F4$\bar{3}$m MgLiAl$_2$ phase is elastically stable while the more-commonly suggested L2$_1$ MgLiAl$_2$ phase is not. 
This may suggest that MgLiAl$_2$ is most-likely only meta-stable and decompose to lower-ordered intermetallics and/or the elemental phases. 
Indeed, the previous work by Levinson and McPherson~\cite{levinson1956phase} have predicted MgLi$_2$Al phase rather than MgLiAl$_2$. 
The both L2$_1$ and F4$\bar{3}$m phases of MgLi$_2$Al exhibit quite promising in terms of strength and hardness.
Similarly, the isolated D0$_3$ Mg$_3$Al does not show any exceptional strength or hardness.
It may be a hint that D0$_3$ Mg$_3$Al leads to exceptional hardening (as suggested by Tang \textit{et al.}~\cite{Tang2019}) due to its semi-coherent precipitation on the Li matrix. 
By doing so, it may lead to exceptional hardening through the precipitation hardening rather than its intrinsic mechanical superiority.  

\subsection{Assessment of hardening mechanisms}
\begin{figure*}[t]
\centering

\includegraphics[width=0.8\textwidth]{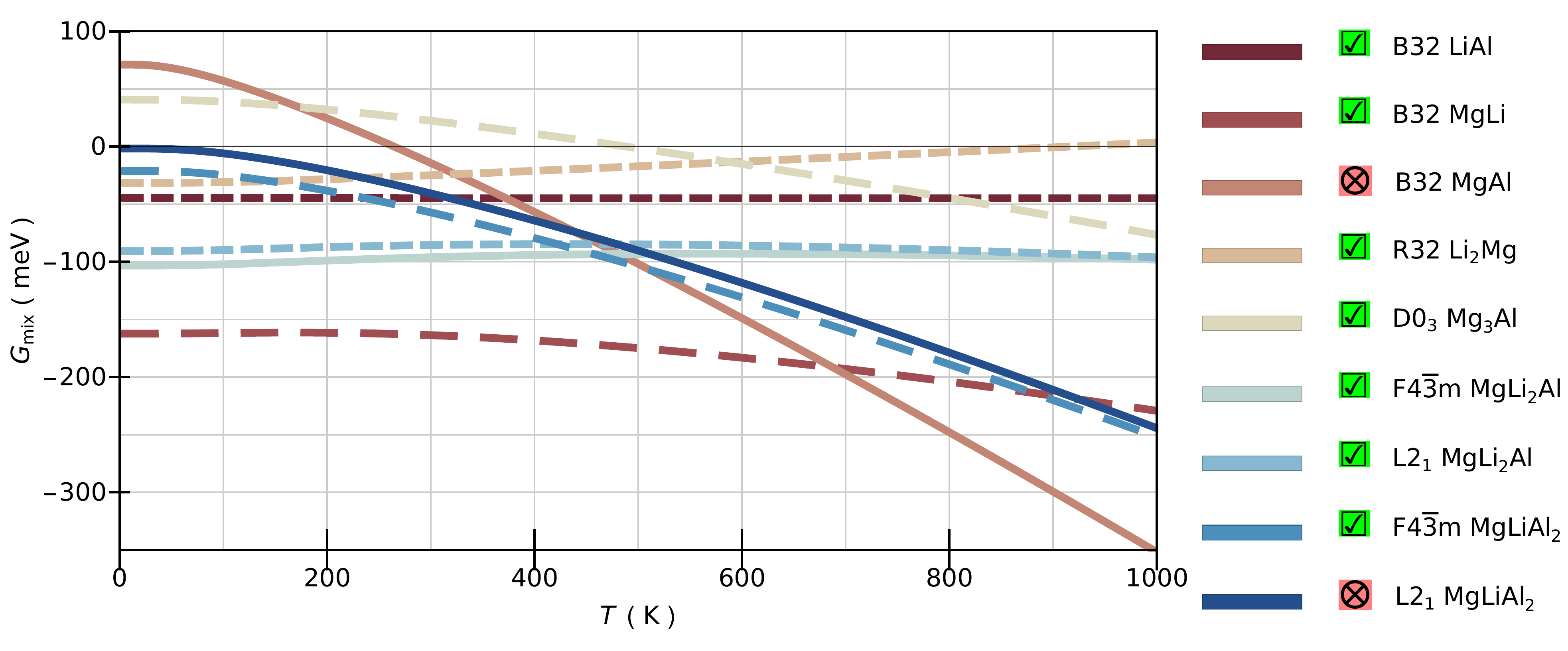}

\caption{Temperature-dependent mixing Gibbs free energies of the most-frequently observed intermetallics in Mg-Li-Al mixtures. 
The elastically stable phases are marked with the green check-box, otherwise with the red corss-box. 
}
\label{fig:Sample-Gmix}
\end{figure*}

We select the subset of intermetallics which have been frequently predicted in the previous experimental works to further gain insight on hardening mechanics. 
In \fig{fig:Sample-Gmix}, the temperature-dependent mixing GFE of the most-frequently appearing phases are shown alongside their elastic-stability test results. 
Among them, the B32 MgAl and D0$_3$ Mg$_3$Al phases have positive $H_\mathrm{F}$ ($G_\mathrm{mix}(T=0) =H_\mathrm{F}$); however, their $G_\mathrm{mix}$ decrease with the increasing temperature. 
B32 MgAl and D0$_3$ Mg$_3$Al become thermally stable at around $\sim 265$~K and $\sim 490$~K, respectively.
On the other hand, the R32 Li$_2$Mg is the only phase with a significantly increasing $G_\mathrm{mix}$ when the temperature rises.
It is thermally stable at the lower temperatures; however, it becomes unstable at around $\sim 920$~K. 
in the legend of \fig{fig:Sample-Gmix}, the results of the elastic-stability tests according to the Born-Huang-stability criteria are shown by the colored boxes for which the green check-box and the red cross-box represent pass and fail, respectively. 
The isolated crystals of the selected phases are elastically stable except the B32 MgAl and L2$_1$ MgLiAl$_2$ phases. 
In particular, the L2$_1$ MgLiAl$_2$ phase is concerning as it is often the assigned symmetry in the XRD measurements. 
However, the F4$\bar{3}$m MgLiAl$_2$ phase is elastically stable with a similar $G_\mathrm{mix}(T)$ to the L2$_1$ MgLiAl$_2$ phase. 
The XRD measurements may not be have enough resolution to distinguish these two symmetries.

\begin{figure}[H]
\centering

\includegraphics[width=0.8\textwidth]{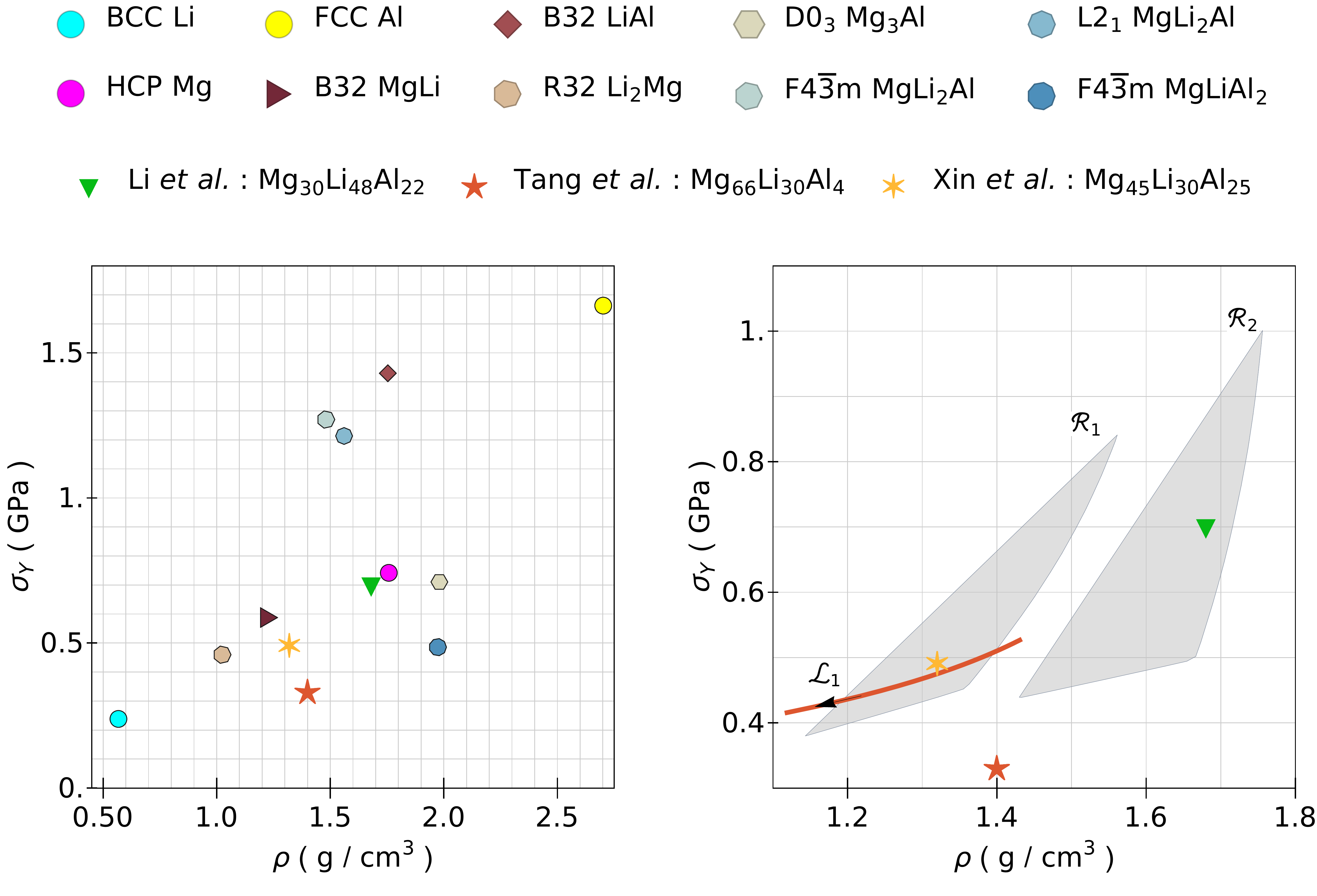}

\caption{Approximate yield strength of the most-frequent intermetallics in the Mg-Li-Al mixtures alongside the experimental values by  Li \textit{et al} for Mg$_{30}$Li$_{48}$Al$_{22}$ \cite{LI2022161703}, Tang \textit{et al.} for Mg$_{66}$Li$_{30}$Al$_{4}$~\cite{Tang2019} and Xin \textit{et al.} for Mg$_{45}$Li$_{30}$Al$_{25}$~\cite{Xineabf3039}. 
Two strength-density regions,  $\mathcal{R}_1$ and $\mathcal{R}_2$ are calculated by applying the rule of mixing to Li, Mg, B32 LiAl, R32 Li$_2$Mg and F4$\bar{3}$m MgLiAl$_2$ (phases observed in the works of Li \textit{et al}) for the MgLiAl and Mg$_{30}$Li$_{48}$Al$_{22}$ compositions.
The strength-density curve, $\mathcal{L}_1$ is obtained by applying the rule of mixing to Li, Mg, D0$_3$ Mg$_3$Al for the Mg$_{70-x}$Li$_{30}$Al$_{x}$. 
}
\label{fig:Sample-Yield}
\end{figure}

In \fig{fig:Sample-Yield} (left), the strength-density profiles of the most-frequently appearing intermetallics in the Mg-Li-Al mixtures are shown alongside the most recent experimental predictions on the hetero-structures. 
Using B32 MgLi, which has $\sigma_\mathrm{y}=0.59$~GPa and $\rho=1.23$~g$\cdot$cm$^{-3}$, as the reference system, the intermetallics with Al contents such as  D0$_3$ Mg$_3$Al and F4$\bar{3}$m MgLiAl$_2$ are not favorable due to significantly higher density with comparable strengths. 
On the other hand, the MgLi$_2$Al intermetallics exhibits remarkable strengths in  between of $\sim 1.2-1.3$~GPa  with a slightly higher density of $\sim 1.5$~g$\cdot$cm$^{-3}$.

In terms of strength, the sample with an average composition of  Mg$_{30}$Li$_{48}$Al$_{22}$ from Ref.\citenum{LI2022161703} is superior at the expense of a higher density compared the samples with average compositions of   Mg$_{66}$Li$_{30}$Al$_{4}$ and Mg$_{45}$Li$_{30}$Al$_{25}$ from Refs.~\citenum{Tang2019}~and~\citenum{Xineabf3039}, respectively. 
In the right panel of \fig{fig:Sample-Yield}, the estimated strength-density regions, $\mathcal{R}_1$ and $\mathcal{R}_2$ are calculated using the rule of mixing for 
\begin{gather*}
\mathcal{R}_1 \; : \; \mathrm{MgLiAl} \\
\mathcal{R}_2 \; : \; \mathrm{Mg}_{30}\mathrm{Li}_{48}\mathrm{Al}_{22} \\
\downarrow \\
c_1 \mathrm{Li} + c_2 \mathrm{Mg} + c_3  \mathrm{LiAl} + c_4 \mathrm{Li}_2\mathrm{Mg} + c_5 \mathrm{MgLiAl}_2
\end{gather*}
Similarly, the strength-density curve, $\mathrm{L}_1$ is calculated for
\begin{align*}
\mathcal{L}_1 \; : \; \mathrm{Mg}_{70- x}\mathrm{Li}_{30}\mathrm{Al}_{x} \rightarrow 30 \mathrm{Li} + (70-4x) \mathrm{Mg} + x \mathrm{Mg}_3\mathrm{Al}.
\end{align*}
The region $\mathrm{R}_1$ represent an hypothetical MgLiAl composition forming the hetero-structures of the intermetallics from Ref.~\citenum{LI2022161703}.
$\mathrm{R}_1$ indicates that the MgLiAl composition may provide significant strengths at reasonably low densities.
The justification of the simple rule of mixing is the region $\mathrm{R}_2$ where the experimentally measured composition of Mg$_{30}$Li$_{48}$Al$_{22}$ is targeted.
The experimental measurement on this composition is located nearly at the center of the estimate strength-density region. 
The curve $\mathrm{L}_1$ is inspired by the work in Refs.~\citenum{Tang2019}~and~\citenum{Xineabf3039} where the effects of increasing Al content and resulting D0$_3$ Mg$_3$Al precipitation has been investigated. 
On remarkable trend is that the increasing Al content indeed reduces the density due to increasing volume fraction of D0$_3$ Mg$_3$Al which has a smaller volume per atom compared to HCP Mg.
The calculate $\mathrm{L}_1$ overestimates the strength of Mg$_{66}$Li$_{30}$Al$_{4}$; however, it predict well the strength of Mg$_{45}$Li$_{30}$Al$_{25}$.
This is quite promising that the simple rule of mixing can be used as a predictive model to estimate the strength-density profiles of Mg-Li-Al hetero-structures starting from the simple Mg-Li-Al intermetallics.
Moreover, the promising hetero-structures can be architectured using the promising intermetallics .

\section{Conclusion}
In this work, we presented a systematic first-principles investigation of formations and strength-density profiles of the Mg-Li-Al LWAs. 
It was expediently shown by applying VCA that the Mg-Li-Al mixtures do not form RSSs. 
Following this, the thermal and elastic stabilities of the BCC-based intermetallics were assessed and the mechanical properties of the elastically stable intermetallics were studied for their possible hardening contributions to the heteros-structures.
Despite its frequent appearance in the experimental measurements on the Mg-Li-Al hetero-structures, MgLiAl$_2$ does not offer any significant hardening compared the conventional Mg-Li alloys such as B32 MgLi.
On the other hand, the MgLi$_2$Al intermetallics can be quite promising hardening source while keeping density low. 
The simple rule of mixing using the isolated intermetallics works remarkable well when estimating the mechanical strength of the complex hetero-structures. 
It was shown that it can be used to  engineer the strength-density profiles of complex structures.   

\section*{Acknowledgment}
We acknowledge the support from the Natural Sciences and Engineering Research Council of Canada (NSERC) through the Discovery Grant under Award Application Number RGPIN-2016-06114, and the New Frontiers in Research Fund (NFRFE-2019-01095).
This research was supported in part through computational resources and services provided by Advanced Research Computing at the University of British Columbia.

\end{document}


\title{{\Large Supplementary Materials}\\``Engineering ultra-strong Mg-Li-Al-based light-weight alloys from first principles"}

\author{Okan K. Orhan}
\author{Mauricio Ponga}
\affiliation{Department of Mechanical Engineering, University of British Columbia, 2054 - 6250 Applied Science Lane, Vancouver, BC, V6T 1Z4, Canada}

\maketitle

\section*{Gibbs free energy}
The Gibbs free energy (GFE) of a non-magnetic, and pristine solid is given by
\begin{align}\label{eq:sm1}
G(V,T)= H(V)+F_\mathrm{el}(V,T)+F_\mathrm{vib}(V,T),
\end{align}
where  $H(V)$ is the formation enthalpy, $F_\mathrm{el}(V,T)$ and $F_\mathrm{vib}(V,T)$ are the electronic and vibrational  Helmholtz free energies, respectively. 

At the equilibrium volume, $H(V)$ is simply  equal to the internal energy. 
Within the approximate KS-DFT, the total energy ($E_0$) is not physically meaningful~\cite{HEINE19701}. 
However, $E_0$ is still useful to approximate the mixing enthalpies when the atomic pseudo-potentials are used consistently. 

The electronic Helmholtz free energy is given for metallic systems by~\cite{WANG20042665,landaystat,10.3389/fmats.2017.00036}
\begin{align}\label{eq:sm2}
F_\mathrm{el}&=
\left(\int dE\; N(E,V)f E - \int^{E_\mathrm{F}} dE\; N(E,V)E\right) \nonumber \\
&+T \left(k_\mathrm{B}\int dE\; N(E,V)\left[f ln(f)+(1-f ) ln(1-f) \right]\right),
\end{align}
where $k_\mathrm{B}$ is the Boltzmann constant, $N(E,V)$ is the electron density of states (DOS), $f=f(E,E_\mathrm{F},T)$ is the Fermi-Dirac distribution function around the Fermi level with the Fermi energy, $E_\mathrm{F}$. 

The simplest approach to obtain the vibrational Helmholtz free energy is within the Debye model, given by ~\cite{anderson1965physical,SHANG20101040}
\begin{align}\label{eq:sm3}
F^\mathrm{D}_\mathrm{vib}=&\frac{9}{8}k_\mathrm{B}\Theta_\mathrm{D}+k_\mathrm{B}T\left\{3\;ln\left[1-e^{-\frac{\Theta_\mathrm{D}}{T}}\right] +D\left[\frac{\Theta_\mathrm{D}}{T}\right]\right\},
\end{align} 
where $\Theta_\mathrm{D}$ is the Debye temperature, and \[D(x)= \frac{3}{x^3} \int_0^x du\; \frac{u^3}{e^u-1}.\]
The Debye temperature is approximately given by~\cite{ANDERSON1963909}
\begin{align}\label{eq:sm4}
\Theta_\mathrm{D}=\frac{h}{k_\mathrm{B}}\left[\frac{3}{4\pi}\frac{N}{V}\right]^{1/3}v_\mathrm{s},
\end{align}
where $N$ is the number of atoms in the unit cell and $v_\mathrm{s}$ is the average sound velocity.
The average sound velocity, $v_s$ can be determined by solving the Christoffel equation in each direction~\cite{JAEKEN2016445}. 

\section*{The Born-Huang-stability criteria for the cubic and hexagonal symmetries}\label{ap:ap2}
The necessary and sufficient conditions for the elastic stability of cubic systems are given by~\cite{PhysRevB.90.224104}

\begin{align}\label{eq:sm5}
c_{11} − c_{12} > 0,\quad c_{11} + 2c_{12} > 0 \; \mathrm{and} \quad c_{44} > 0
\end{align}
where $c_{ij}$ are the elements of the second-order elastic tensor  $\mathbb{C}$.

In the case of the hexagonal symmetry, the  necessary and sufficient conditions are
\begin{align}\label{eq:sm6}
&c_{11} > |c_{12}|,\quad  c_{33}\left(c_{11}+c_{12}\right) > 2c_{13}^2 \; \mathrm{and} \quad  c_{44} >0 .
\end{align}

In the case of the rhombohedral symmetry, the  necessary and sufficient conditions are
\begin{align}\label{eq:sm7}
& c_{11} > |c_{12}|,\quad  c_{33}\left(c_{11}+c_{12}\right) > 2c_{13}^2 ,\nonumber \\
& c_{44}c_{66} > c_{14}^2 \; \mathrm{and} \quad  c_{44} >0 .
\end{align}

\section*{Elastic constants of the cubic and hexagonal systems}
The Voigt-Reuss-Hill (VHR)~\cite{Hill_1952} averaged bulk ($B$) and shear ($S$) modulus are given by
\begin{align}\label{eq:sm8}
B_\mathrm{VHR}=\frac{B_\mathrm{V}+B_\mathrm{R}}{2},\quad \mathrm{and}\quad S_\mathrm{VHR}=\frac{S_\mathrm{V}+S_\mathrm{R}}{2},
\end{align}
where the sub-indices V and R represent the Voigt and Reuss averagings~\cite{doi:10.1002/andp.18892741206,doi:10.1002/zamm.19290090104} which provide the upper and lower bounds, respectively.
The Voigt- and Reuss-averaged bulk and shear modulus for cubic systems are given by~\cite{HAO20203488}
\begin{align}\label{eq:sm9}
B_\mathrm{V}&=B_\mathrm{R}=\frac{c_{11}+2c_{12}}{3}, \nonumber \\
S_\mathrm{V}&=\frac{c_{11}-c_{12}+3c_{44}}{5},\quad \mathrm{and}\quad  S_\mathrm{R}=\frac{5(c_{11}-c_{12})c_{44}}{3c_{11}-3c_{12}+4c_{44}}.
\end{align}
Similarly, the Voigt- and Ruess-averaged bulk and shear modulus for hexagonal systems are given by~\cite{tromans2011elastic}
\begin{align}
B_\mathrm{V}&=\frac{2(c_{11}+c_{12}+0.5 c_{33}+2c_{13}}{9},\quad \mathrm{and}\quad B_\mathrm{R}=\frac{1}{3\alpha+6\beta} \nonumber \\
S_\mathrm{V}&=\frac{7c_{11}-5c_{12}+12c_{44}+2c_{33}-4c_{13}}{30},\quad \mathrm{and} \nonumber \\ S_\mathrm{R}&=\frac{5}{4\alpha-4\beta+3\lambda},
\end{align}
where $3\alpha=s_{11}+s_{11}+s_{33}$, $3\beta=s_{23}+s_{31}+s_{12}$, and $3\lambda=s_{44}+s_{55}+s_{66}$.
$s_{ij}$ is elements of the elastic compliances matrix given by $\mathbb{S}=\mathbb{C}^{-1}$.
Then, Young's modulus ($Y$) and the Poisson ratio ($\nu_\mathrm{P}$)  are given in terms of $B$ and $S$ by
\begin{align}
Y =\frac{9 B S}{3B + S}, \quad \mathrm{and}\quad \nu_\mathrm{P}=\frac{3B-2S}{6B+2S}. 
\end{align}

\section*{Computational details}
The SG15 optimized norm-conserving Vanderbilt scalar-relativistic pseudo-potentials~\cite{PhysRevB.88.085117,SCHLIPF201536} using  the Perdew-Burke-Ernzerhof  exchange-correlation functional~\cite{PhysRevLett.43.1494,Kerker_1980,PhysRevB.40.2980} were used for the Mg, Li, and Al atomic pseudo-potentials.
Initial crystallographic information of the stable phases of Mg, Li, and Al were extracted from Ref.~\citenum{doi:10.1063/1.4812323}.
Initial crystallographic information were obtained by substituting the virtual atoms for Mg, Li, and Al during random solid solution simulations at the disordered mean-field limit. 
The initial crystal structures for the available body-centered cubic (BCC) intermetallics were constructed by using the generic templates, shown in Fig.~\ref{fig:BCC-based-Str}.

Variable-cell geometry optimizations were performed using the Quantum ESPRESSO (QE) software~\cite{0953-8984-21-39-395502,0953-8984-29-46-465901} with a $150$~Ry kinetic-energy cutoff, a  Marzari-Vanderbilt cold-smearing~\cite{PhysRevLett.82.3296} equivalent to  $300$~K, and a $12\times12\times12$ Monkhorst-Pack-equivalent~\cite{PhysRevB.13.5188} uniform Brillouin zone with a $10^{-7}$~Ry total-energy, and $10^{-6}$~Ry/a$_0$ total force convergence and $10^{-10}$ self-consistency thresholds. 
Thermodynamic quantities, the elastic properties, and the phonon dispersion were calculated using the  \textit{"thermo\_pw"} package~\cite{ThermoPW} within the QE software with a commonly converged kinetic-energy cutoff of $100$~Ry on a $12\times12\times12$ shifted Monkhorst-Pack-equivalent uniform Brillouin zone.

\begin{figure}[H]
\centering

\subfloat[Elemental metals]{
\includegraphics[width=0.2\textwidth]{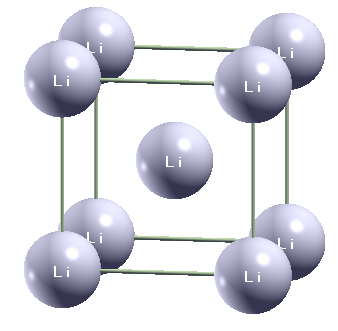}\hspace{1.5cm}
\includegraphics[width=0.2\textwidth]{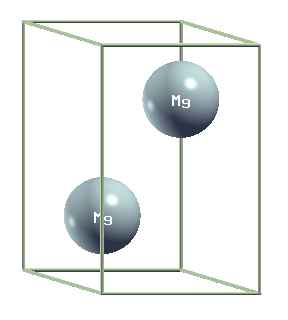}\hspace{1.5cm}
\includegraphics[width=0.2\textwidth]{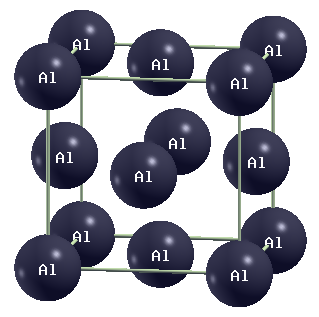}}

\subfloat[B2 ordered structure $AB$ alloys]{
\includegraphics[width=0.3\textwidth]{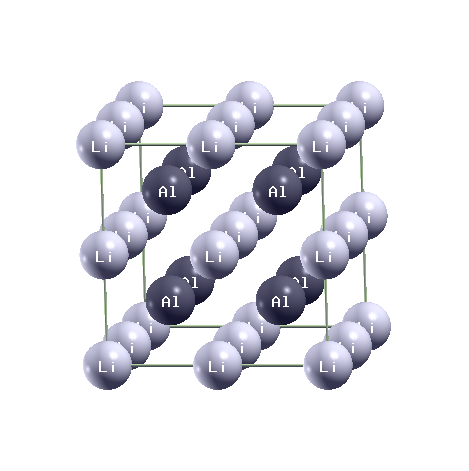}
\includegraphics[width=0.3\textwidth]{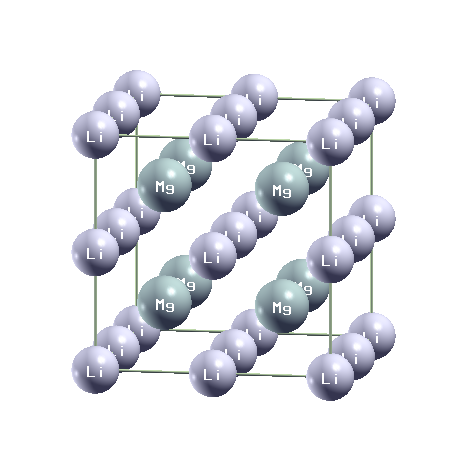}
\includegraphics[width=0.3\textwidth]{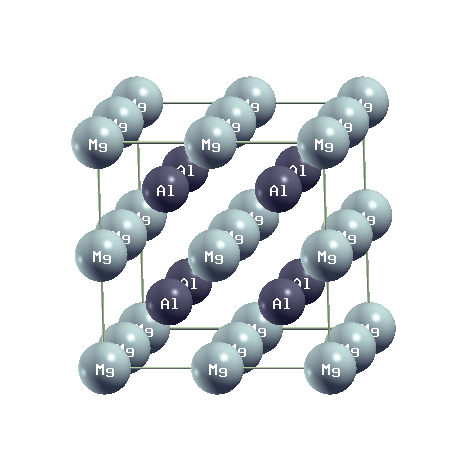}}

\subfloat[B32 ordered structure $AB$ alloys]{
\includegraphics[width=0.3\textwidth]{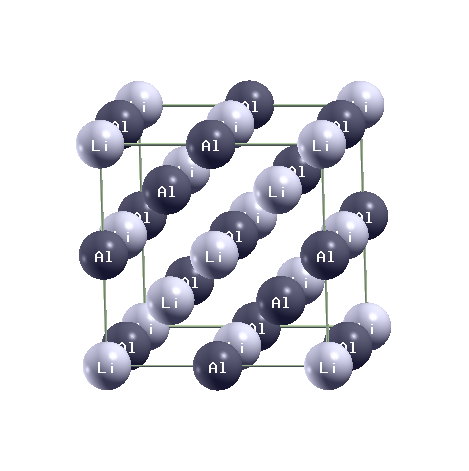}
\includegraphics[width=0.3\textwidth]{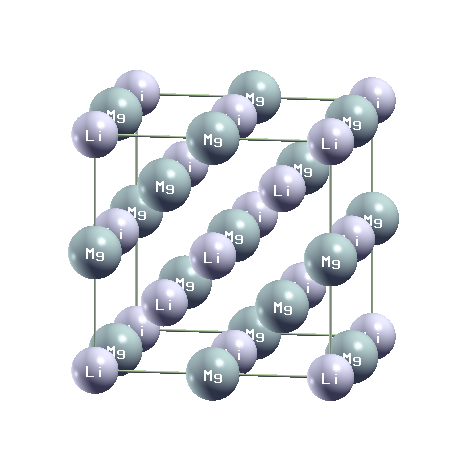}
\includegraphics[width=0.3\textwidth]{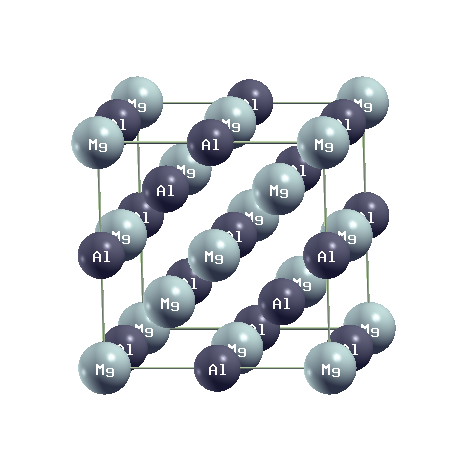}}

\end{figure}

\begin{figure}[H]
\centering
\ContinuedFloat

\subfloat[D0$_3$ ordered structure $A_3B$ alloys]{
	\begin{tabular}[b]{ccc}
		\includegraphics[width=0.3\textwidth]{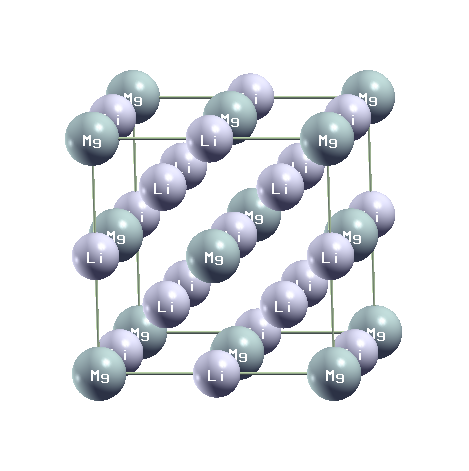} &
		\includegraphics[width=0.3\textwidth]{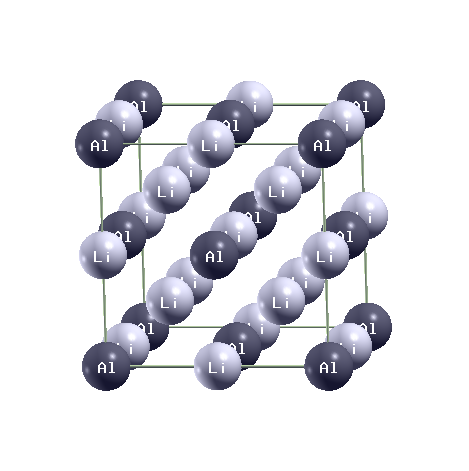} &
		\includegraphics[width=0.3\textwidth]{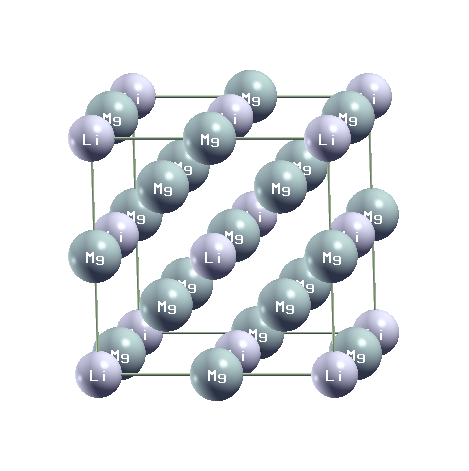}\\  
		\includegraphics[width=0.3\textwidth]{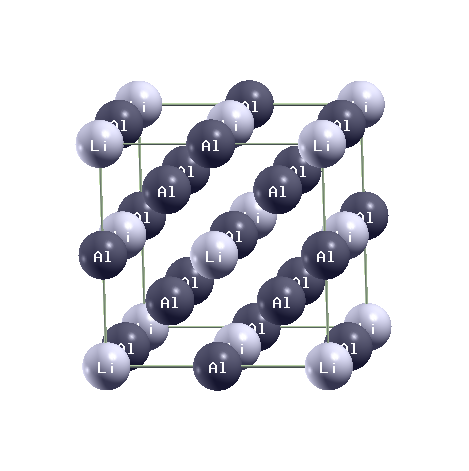}&
		\includegraphics[width=0.3\textwidth]{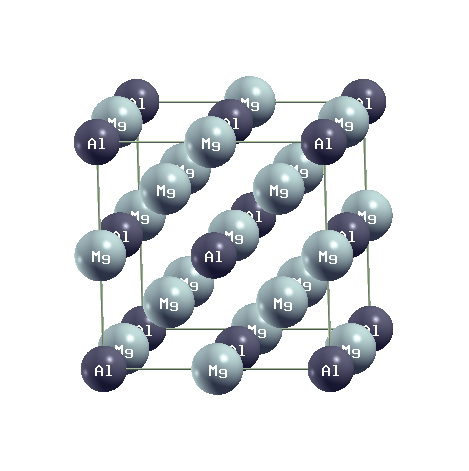}&
		\includegraphics[width=0.3\textwidth]{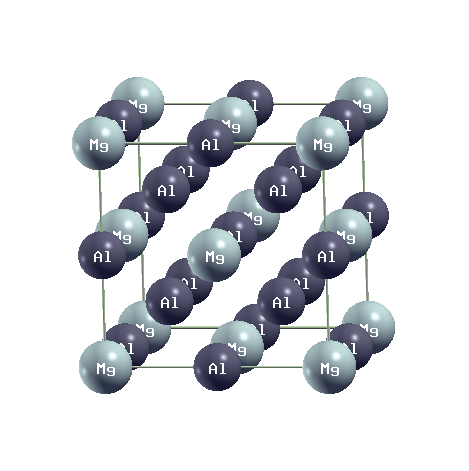}          
	\end{tabular}}
	
\subfloat[F4$\bar{3}m$ ordered structure $A_2BC$ alloys]{
\includegraphics[width=0.3\textwidth]{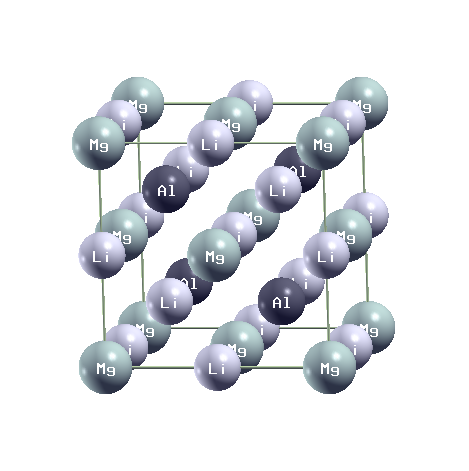}
\includegraphics[width=0.3\textwidth]{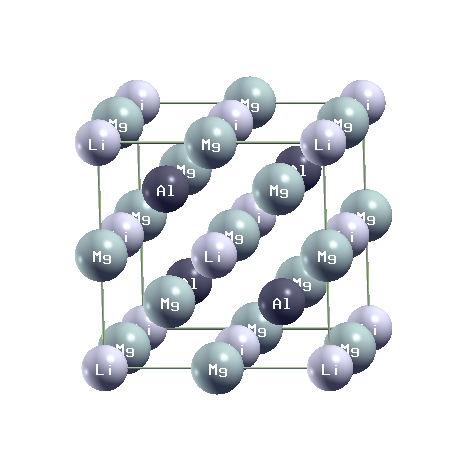}
\includegraphics[width=0.3\textwidth]{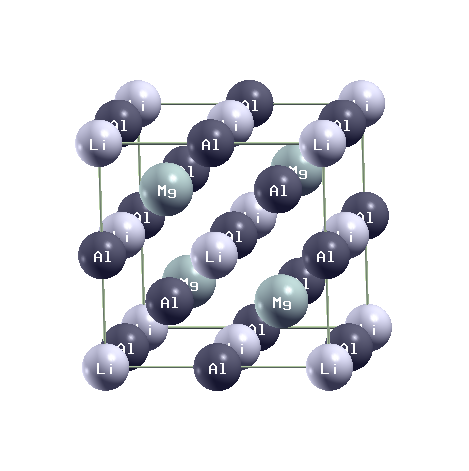}}

\end{figure}

\begin{figure}[H]
\centering
\ContinuedFloat

\subfloat[F4$\bar{3}m$ ordered structure $A_2BC$ alloys]{
\includegraphics[width=0.3\textwidth]{F4b3M-Li2MgAl}
\includegraphics[width=0.3\textwidth]{F4b3M-LiMg2Al}
\includegraphics[width=0.3\textwidth]{F4b3M-LiMgAl2}}

\subfloat[R32 Li$_2$Mg]{
\includegraphics[width=0.17\textwidth]{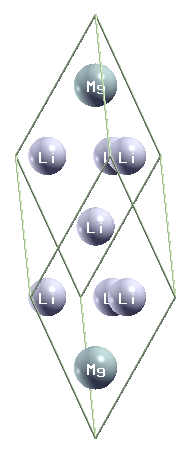}}

\caption{Crystal structures of Mg, Li, and Al intermetallics.}
\label{fig:BCC-based-Str}
\end{figure}

\section*{Temperature-dependence of free energies of the body-centered-cubic intermetallics}

\begin{figure}[H]
\centering

\includegraphics[width=0.8\textwidth]{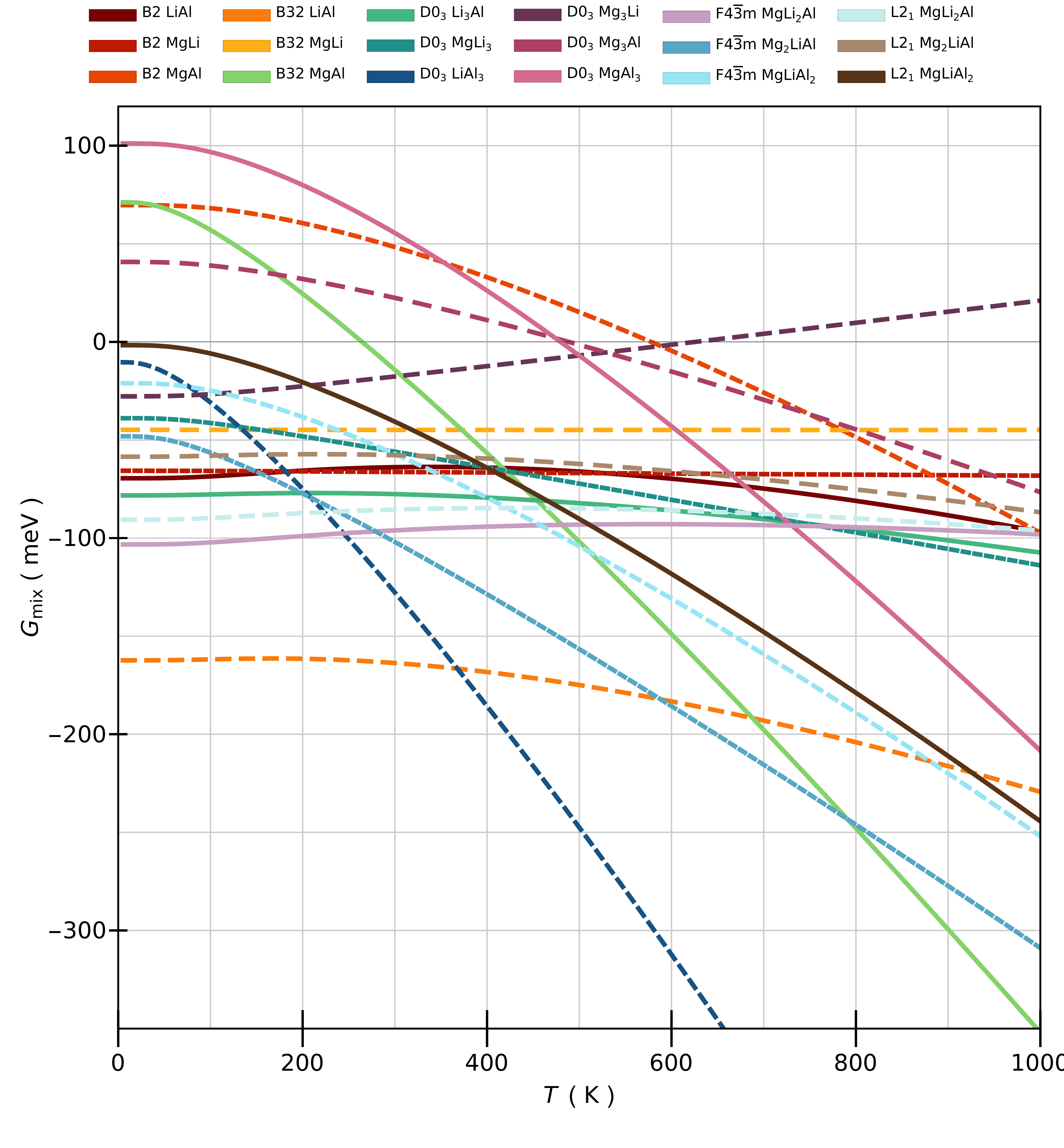}

\caption{Temperature dependence of the mixing Gibbs free energies of the body-centered-cubic-based intermetallic of  Mg, Li and Al.
The predominant terms are the vibrational Helmholtz energies in determining the temperature-dependent trends.
The vibrational Helmholtz energies are calculated within the Debye model.
}
\label{fig:BCC-Gmix-TDep}
\end{figure}

\begin{figure}[H]
\centering

\includegraphics[width=1.0\textwidth]{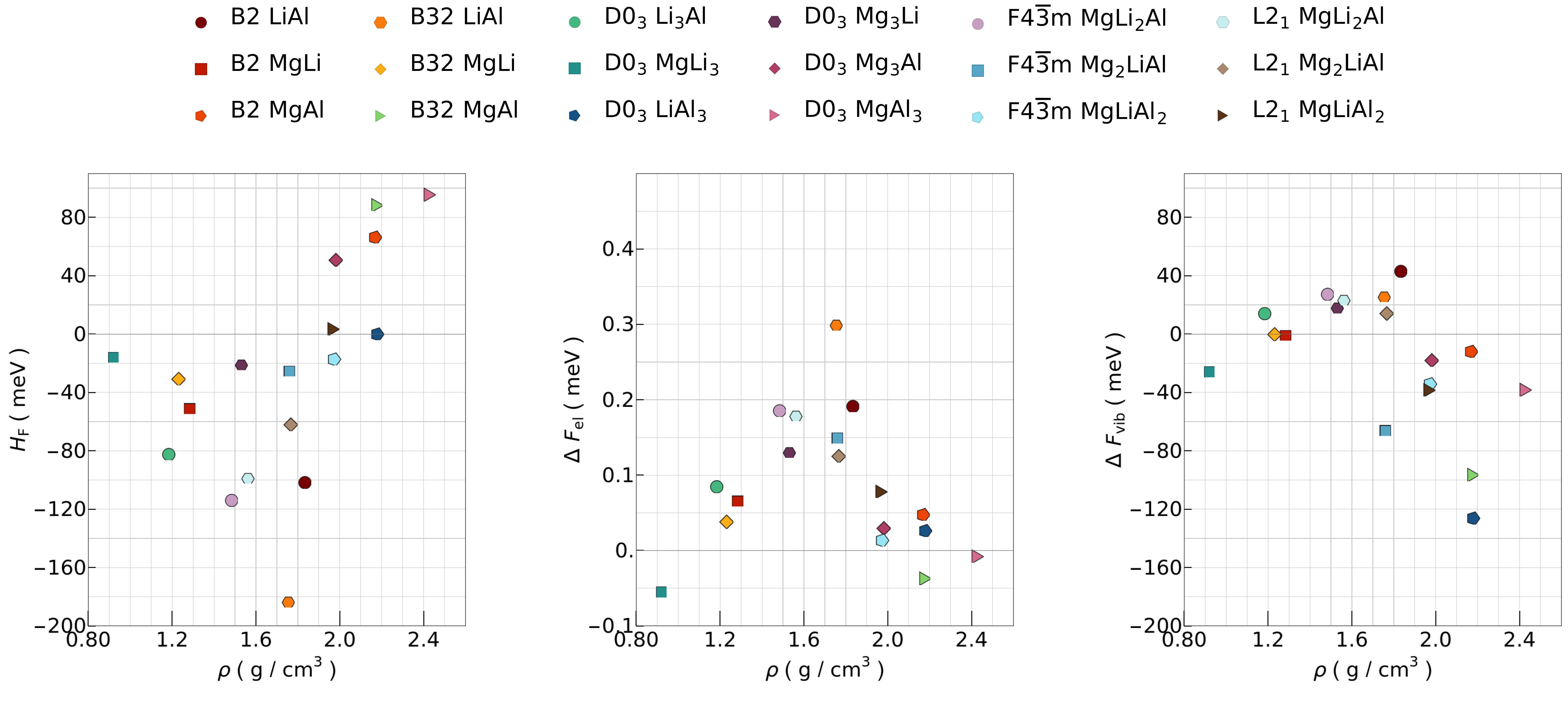}

\caption{The three main contribution to the mixing Gibbs free energies at $300$~K.}
\label{fig:BCC-HoF}
\end{figure}